\documentclass[final]{svjour3}
\pdfoutput=1
\usepackage{graphicx}
\usepackage{rotating}
\usepackage{amssymb}
\usepackage{mathptmx}
\usepackage[numbers]{natbib}

\usepackage{amsmath}
\usepackage{bm} 
\usepackage[version=3]{mhchem}
\usepackage[colorlinks]{hyperref}
\usepackage[caption=false]{subfig}

\makeatletter
\journalname{Journal of Low Temperature Physics}

\bibpunct{[}{]}{,}{n}{}{,}

\begin{document}

\newcommand{\hdblarrow}{H\makebox[0.9ex][l]{$\downdownarrows$}-}
\title{Calculation of Leggett-Takagi relaxation in vortices of superfluid \ce{^{3}He}-B}

\author{S. M. Laine \and E. V. Thuneberg}

\institute{Department of Physics, University of Oulu, Finland }

\maketitle

\begin{abstract}

We calculate the relaxation of Brinkman-Smith mode via Leggett-Takagi relaxation in the presence of an isolated vortex in superfluid \ce{^{3}He}-B. The calculation is based on an  analytical solution of the order parameter far from the vortex axis. We obtain an expression for the dissipated power per vortex length as a function of the tipping angle of the magnetization and the orientation of the static magnetic field with respect to the vortex.


\end{abstract}

\section{Introduction}
Vortices of two different core structures have been observed in superfluid  \ce{^{3}He}-B \cite{Ikkala82}. The vortex that is stable in the major part (low pressure and temperature) of the phase diagram was theoretically identified as the double-core vortex. The vortex that is stable in the high pressure and temperature corner of the superfluid phase diagram was identified as the A-phase-core vortex \cite{thuneberg1987}.
The main experimental tool to study the vortices has been nuclear magnetic resonance (NMR).
Two distinct NMR modes have been used. For small tipping of the magnetization one can see the collective effect of vortices on the texture via frequency shifts of spin wave modes \cite{Hakonen89}. For larger tipping one gets the Brinkman-Smith mode \cite{Brinkman75b}, where the texture is erased. At angles exceeding the Leggett angle $104^\circ$ this forms the homogeneously precessing domain (HPD) \cite{kondo1991}. Here the spins within the domain precess uniformly even in a nonuniform magnetic field.  The measured quantity is the absorbed energy. Based on the relative jump of the observables at the transition between the vortex types, the latter method is more sensitive to the vortex structure than the former \cite{kondo1991}.

Several mechanisms contribute to the absorption in the Brinkman-Smith mode. In this paper we study theoretically one of these, the Leggett-Takagi relaxation \cite{leggett1977}.
In the model developed by Leggett and Takagi, the magnetization of superfluid \ce{^{3}He} is comprised of two parts, a superfluid (Cooper pair) component and a normal (quasiparticle) component. As the dipole force acts only on the superfluid component, a nonequilibrium arises between the components. Dissipation arises as the components relax towards their mutual equilibrium values. In vortices the major part of the relaxation arises from the region outside the vortex core. Our calculation uses an analytic solution of the order parameter in this region. We compare our results with the experiment in Ref.\ \cite{kondo1991}. While there is a partial agreement, it seems that Leggett-Takagi relaxation alone is insufficient to explain the measurements.

\section{Asymptotic structure of B-phase vortices}

The order parameter of an isolated B-phase vortex far from the vortex axis can be written as
\begin{equation}
\mathsf{A} = e^{i \varphi} \Delta_0 \mathsf{R} \left( \theta_0 \hat{\bm{n}}  \right) \mathsf{R} \left( \bm \theta  \right).
\end{equation}
Here $\varphi$ is the azimuthal angle around the vortex axis, $\Delta_0$ is the bulk gap,  $\mathsf{R} \left( \theta_0 \bm{\hat{n}}  \right)$ is a rotation by the Leggett angle $\theta_0=\arccos(-1/4)\approx 104^{\circ}$ around an axis $\hat{\bm{n}}$, determined by the bulk, and $\mathsf{R} \left( \bm \theta  \right)$ is an additional rotation by an angle $ \theta = \left| \bm \theta  \right|$ around the axis $ \hat{\bm\theta} = \bm \theta / \theta$. In the static situation $\bm \theta$ is determined by minimizing the free energy \cite{vollhardt1990, thuneberg2001}
\begin{equation}\label{eq:FreeEnergy}
F = \int d^3 r \left(f_D + f_G  \right).
\end{equation}
Here $f_D$ originates from the dipole-dipole interaction between the \ce{^{3}He} nuclei. Neglecting constants, it has the expression
\begin{equation} f_{D}=\lambda_{\rm D}(R_{ii}R_{jj}+R_{ij}R_{ji})
=\textstyle{\frac{1}2}\lambda_{\rm D}(4\cos\vartheta+1)^2
\label{eq:dipegen}\end{equation}
where $\mathsf{R}(\bm\vartheta)=\mathsf{R} \left( \theta_0 \hat{\bm{n}}  \right) \mathsf{R}\left( \bm \theta  \right)$ is the total spin-orbit rotation.
For small $\bm\theta$ we can approximate
\begin{equation}\label{eq:dipea} f_{D}
=\textstyle{\frac{15}2}\lambda_{D}
(\hat{\bm n}\cdot\bm \theta)^2
\end{equation}
with an error proportional to $\theta^3$.
The gradient term $f_G$ to quadratic order in derivatives is
\begin{equation}\label{eq:graded}
f_G = 2 \lambda_{G2} \left[ (1 + c) \partial_i \theta_k  \partial_i \theta_k - c \partial_i \theta_k  \partial_k \theta_i \right],
\end{equation}
where $c = \lambda_{G1} / 2 \lambda_{G2}$. The gradient energy would be simplified if $c$ vanished. However, $c=1$ is expected to be closer to reality as this value is obtained in the weak-coupling approximation neglecting Fermi-liquid parameters $F_1^a$ and $F_3^a$. 

An important quantity in the following is the torque acting on the superfluid magnetization. It is given by the variational derivative
\begin{equation}\label{eq:torque}
\bm R=-\frac{\delta F}{\delta\bm\theta}=-15\lambda_D \hat{\bm n}(\hat{\bm n}\cdot\bm\theta)+4\lambda_{G2}\left[(1+c)\nabla^2\bm \theta-c\bm\nabla(\bm\nabla\cdot\bm\theta)\right].
\end{equation}
Here the latter equality follows from expressions (\ref{eq:FreeEnergy}), (\ref{eq:dipea}) and (\ref{eq:graded}). In equilibrium, the order parameter is determined by the minimum of the free energy. This condition leads to the fact that the torque (\ref{eq:torque}) has to vanish,
\begin{equation}
\bm R \equiv 0.
\label{e.difeqt}\end{equation}
At distances much less than the dipole length $\xi_D = ( \lambda_{G2} / \lambda_D)^{1/2}$, we can neglect the dipole term ($\lambda_D\rightarrow 0$). In this case the solution appropriate for an isolated vortex is
\begin{equation}\label{eq:theta}
\bm \theta = \frac{C_1 \cos \varphi}{r} \left( \frac{\sin \varphi}{1+c} \bm{\hat{r}} + \cos \varphi \bm{\hat{\varphi}}  \right) + \frac{C_2 \sin \varphi}{r} \left( - \frac{\cos \varphi}{1+c} \bm{\hat{r}} + \sin \varphi \bm{\hat{\varphi}}  \right).
\end{equation}
Here ${\hat{\bm r}}$, $\bm{\hat{\varphi}}$ and $\bm{\hat{z}}$ are the basis vectors of the cylindrical coordinate system, with $\bm{\hat{z}}$ oriented along the vortex axis. 
The temperature- and pressure-dependent constants $C_1$ and $C_2$ determine the type of the vortex. They can be obtained from the numerical solution of the vortex core structure \cite{thuneberg1987,fogelstrom1995, silaev2015}. For the A-phase-core vortex $C_1 = C_2$. This special  case  of Eq. (\ref{eq:theta}), $\bm \theta=C_1\bm{\hat{\varphi}}/r$,  was found by Hasegawa \cite{Hasegawa85}. For the double-core vortex the coefficients differ essentially, $C_1 / C_2 \gg 1$. The inclusion of the dipole term in (\ref{e.difeqt}) causes $\bm \theta$ to vanish more rapidly than the $r^{-1}$ dependence in (\ref{eq:theta}) at distances greater than $\xi_D$. This will be discussed in a moment.

We study NMR in the presence of a static magnetic field $\bm B$. In the Brinkman-Smith mode, the unit vector $\bm{\hat{n}}$ precesses uniformly around $\bm B$ at an angular velocity $\omega$. If the tipping angle of the magnetization, $\beta$, measured from the direction of $\bm B$, is less than the Leggett angle $\theta_0$, the precession is at the Larmor frequency, $\omega = \omega_L = \gamma_0 B$, and $(\hat{\bm n}\cdot\hat{\bm B})^2=(4\cos\beta+1)/5$. If $\beta$ is larger than $\theta_0$,  the precession rate is  increased and $\hat{\bm n}$ stays perpendicular to ${\bm B}$ \cite{Brinkman75b}. Under usual experimental conditions the vortices do not follow the precession of $\bm{\hat{n}}$ \cite{kondo1991}. We assume that the spin-orbit rotation field $\bm\theta(\bm r)$ takes a time-independent value in the Brinkman-Smith mode. This field  is determined by vanishing of the {\em time-averaged} torque,
\begin{equation}\label{eq:torqueav}
\langle\bm R\rangle=-15\lambda_D \langle\hat{\bm n}\hat{\bm n}\rangle\cdot\bm\theta+4\lambda_{G2}\left[(1+c)\nabla^2\bm \theta-c\bm\nabla(\bm\nabla\cdot\bm\theta)\right]=0.
\end{equation}
This equation can be solved as follows. We first split the solution into two parts, $\bm \theta = \bm \theta_1 + \bm \theta_2$, where $\bm \theta_1$ is given by Eq. (\ref{eq:theta}). This way we force the correct behaviour of $\bm \theta$ near the vortex core. Since the equation is linear, we transform to Fourier space, where the solution is simple. Taking the inverse Fourier transform, $\bm \theta$ can be written in form
\begin{equation}\label{eq:theta_with_dipole}
\bm \theta (\bm r) = \sum_{n = 0}^\infty \left\{ \cos \left[ (2 n + 1) \varphi  \right] \bm{c}_n (r) + \sin \left[ (2 n + 1) \varphi  \right] \bm{s}_n (r) \right\},
\end{equation}
where the coefficients $\bm{c}_n(r)$ and $\bm{s}_n(r)$ are given by
\begin{align}
\bm{c}_n (r) &= \frac{i}{2 \pi^2} \left( -1 \right)^n \int_0^{2 \pi} d \varphi_k \cos \left[ (2 n + 1) \varphi_k \right] \int_0^\infty d k k  J_{2 n + 1}(k r)  \bm \theta (k, \varphi_k) \label{eq:theta_coeff_c} \\
\bm{s}_n (r) &= \frac{i}{2 \pi^2} \left( -1 \right)^n \int_0^{2 \pi} d \varphi_k \sin \left[ (2 n + 1) \varphi_k \right] \int_0^\infty d k k  J_{2 n + 1}(k r)  \bm \theta (k, \varphi_k) \label{eq:theta_coeff_s}.
\end{align}
Here $J_{2 n + 1}(k r)$ is a Bessel function of the first kind of order $2n + 1$ and $\bm \theta (k, \varphi_k)$ is the Fourier transform of $\bm \theta(\bm r)$, written in terms of the $\bm k$-space polar coordinates $\left( k, \varphi_k \right)$. It turns out that we don't have to solve these coefficients explicitly in order to calculate the dissipation, and so we won't do it here. We can, however, describe them qualitatively. For $r \ll \xi_D$ the coefficients $\bm{c}_n(r)$ and $\bm{s}_n(r)$ behave like $r^{-1}$ for $n = 0, 1$, but remain finite for $n \geq 2$. For $r \gg  \xi_D$, $\bm{c}_n(r)$ and $\bm{s}_n(r)$ vanish exponentially.

\section{Leggett-Takagi relaxation}

In a dynamical state the torque $\bm R$ (\ref{eq:torque}) does not vanish in general. This shifts the balance of normal and superfluid components of magnetization, as the the torque is applied only on the latter component. The relaxation of the two components towards their mutual equilibrium leads to dissipation. For energy rate of change Leggett and Takagi derived the formula \cite{leggett1977} 
\begin{equation}\label{eq:dEdt_LT}
\left( \frac{d E}{d t}  \right)_{LT} = - \frac{\mu_0 \gamma_0^2}{\chi_0} \frac{1 - \lambda}{\lambda} \tau \int d^3 r \left| \bm R  \right|^2.
\end{equation}
Here $\mu_0$ is the vacuum permeability, $\gamma_0$ the gyromagnetic ratio, $\chi$ the magnetic susceptibility, $\chi_0$ the magnetic susceptibility in the absence of the Fermi liquid effects, $\lambda$ the equilibrium fraction of the
superfluid magnetization, and $\tau$ the Leggett-Takagi relaxation time.
In order to find the averaged energy dissipation, we have to evaluate the average $\langle |\bm R|^2\rangle$. Substituting $\bm R$ from Eq.\ (\ref{eq:torque}) and using Eq.\ (\ref{eq:torqueav}) we get
\begin{equation}\label{eq:R2}
\langle\left| \bm R  \right|^2\rangle =  (15\lambda_D)^2\bm \theta\cdot\left(\langle\bm{\hat{n}}\bm{\hat{n}}\rangle-\langle\bm{\hat{n}}\bm{\hat{n}}\rangle\cdot\langle\bm{\hat{n}}\bm{\hat{n}}\rangle\right)\cdot\bm\theta.
\end{equation}
Here $\bm \theta$ is given by Eq. (\ref{eq:theta_with_dipole}). Since the coefficients $\bm{c}_n(r)$ and $\bm{s}_n(r)$ diverge as $r^{-1}$ near the origin, we need to introduce a cut-off radius $r_1 \ll  \xi_D$ in order for the integral in (\ref{eq:dEdt_LT}) to converge. The value of $r_1$ will be discussed later. 
Keeping only the leading term in $r_1$, the time-averaged dissipated power per vortex length is
\begin{equation}\label{eq:P_LT}
P_{LT} = - \frac{\chi \Omega^4}{\mu_0 \gamma_0^2} \widetilde{\tau}_{LT} \ln\left( \sqrt{\frac4{15}}\frac{\xi_D}{r_1} \right)  \widetilde{P}.
\end{equation}
Here
\begin{equation}
\widetilde{\tau}_{LT} = \frac{\chi}{\chi_0} \frac{1 - \lambda}{\lambda} \tau
\end{equation}
is the effective relaxation time and $\Omega = (15 \mu_0 \gamma_0^2 \lambda_{D} / \chi)^{1/2}$  the longitudinal resonance frequency. The angular dependence of the power is contained in  
\begin{equation}
\begin{split}
\widetilde{P} &= \frac{\pi \left( 1 - \cos \beta  \right)}{100 (1 + c)^2 } \Bigg\{ \Big[ \big(4 + 6 c +3c^2 \big) C_1^2 + 2 c (2 + c) C_1 C_2 + \big(4 + 6 c +3c^2  \big) C_2^2  \Big] \\
& \times \Big[ 6 + 4 \cos \beta - \big( 1 - 6 \cos \beta  \big) \sin ^2 \eta  \Big] \\
& + 2 (1 + c) (2 + c) \left( C_1^2 - C_2^2  \right) \big( 1 - 6 \cos \beta  \big) \cos 2 \xi \sin^2 \eta \Bigg\}.
\end{split}
\end{equation}
The above expression for $P_{LT}$ holds when $\beta \leq \theta_0$. If the magnetization is tipped more than this, there will be extra dissipation coming from the bulk, but the contribution coming from the vortices will no longer depend on $\beta$, $P_{LT} \left( \beta > \theta_0 \right) = P_{LT} \left( \beta = \theta_0 \right)$. 
The geometry of the system is such that the $z$-axis is along the vortex axis.
The angle $\xi$ describes the orientation of the vortex in the plane perpendicular to the vortex axis, i.e. the $xy$-plane. The $x$-axis and the $y$-axis are chosen so that the anisotropy vector $\bm{\hat{b}}$ of the double-core vortex, pointing from one of the half cores to the other, is given by $\bm{\hat{b}} = -\sin \xi \bm{\hat{x}} + \cos \xi \bm{\hat{y}}$. 
The angle $\eta$ is the tilting angle of $\bm B$ from the vortex axis, $\bm B = B ( \bm{\hat{z}} \cos \eta + \bm{\hat{x}} \sin \eta)$.
Together the angles $\eta$ and $\xi$ completely determine the relative orientation of the vortex with respect to the magnetic field. Note that $\xi$ is only relevant for the double-core vortex, since the A-phase-core vortex is symmetric.

In the following we analyze some special cases of $P_{LT}$.  For numerical values we use $C_1$ and $C_2$ taken from calculations in Ref.\ \cite{silaev2015}, see Table \ref{tab:parameters}. These coefficients are calculated for the conditions of Ref.\ \cite{kondo1991}: pressure 29.3 bar and temperature  $T \sim 0.5T_c$. The coefficients are proportional to the length scale $R_0 = (1 + F_1^s / 3) \xi_0$, where $F_1^s$ is a Fermi liquid parameter, $\xi_0 = \hbar v_F / 2 \pi k_B T_c$ the coherence length, $v_F$ the Fermi velocity, and $T_c$ the critical temperature. Consistently with Ref.\ \cite{silaev2015}, we also assume $c=1$.

\begin{figure}[tb]
\begin{center}
\includegraphics[width=0.6\textwidth]{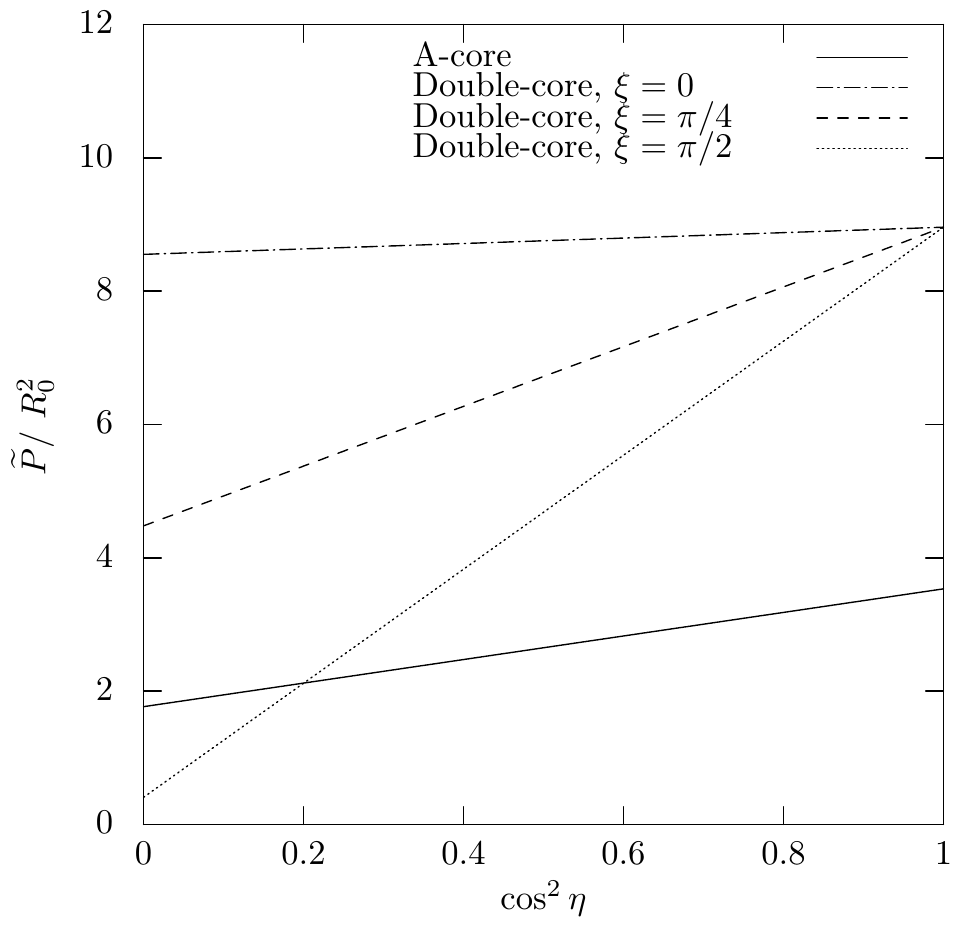}
\end{center}
\caption{$\widetilde{P} / R_0^2$ as a function of $\cos^2 \eta$ at $T = 0.5 T_c$, $p = 29.3$ bar, $\beta = \theta_0$, $c = 1$. $\widetilde{P}$ is independent of $\xi$ for the A-phase-core vortex, but there is a strong dependence on $\xi$ for the double-core vortex when the magnetic field is tilted away from the vortex axis.}
\label{P_tilted}
\end{figure}

If $\beta$ and $\xi$ are fixed, we can write $\widetilde{P} = a_0 + a_2 \cos^2 \eta$, similarly as in Ref. \cite{kondo1991}.
Fig.\ \ref{P_tilted} shows $\widetilde{P}$ plotted as a function of $\cos^2 \eta$ in units of $R_0^2$. The tipping angle is fixed at $\beta = \theta_0$. The result for the A-phase-core vortex does not depend on $\xi$, and $a_2 / a_0 = 1$, independent of temperature and  pressure. The measured value reported in Ref.\ \cite{kondo1991} is $a_2 / a_0 \approx 1.02$. The result for the double-core vortex, on the other hand, is highly dependent on $\xi$. The ratio $a_2 / a_0$ ranges from $0.05$ at $\xi = 0$ to $21.4$ at $\xi = \pi / 2$, with the measured value of $a_2 / a_0 \approx 4.87$. The susceptibility anisotropy of the double-core vortex favors the orientation $\xi=\pi/2$ in tilted field \cite{thuneberg1987}. Based on this, the lowest dissipation as a function of $\xi$ should be expected in Fig.\ \ref{P_tilted}.

\begin{figure}[tb]
\subfloat{%
\includegraphics[width=0.49\textwidth]{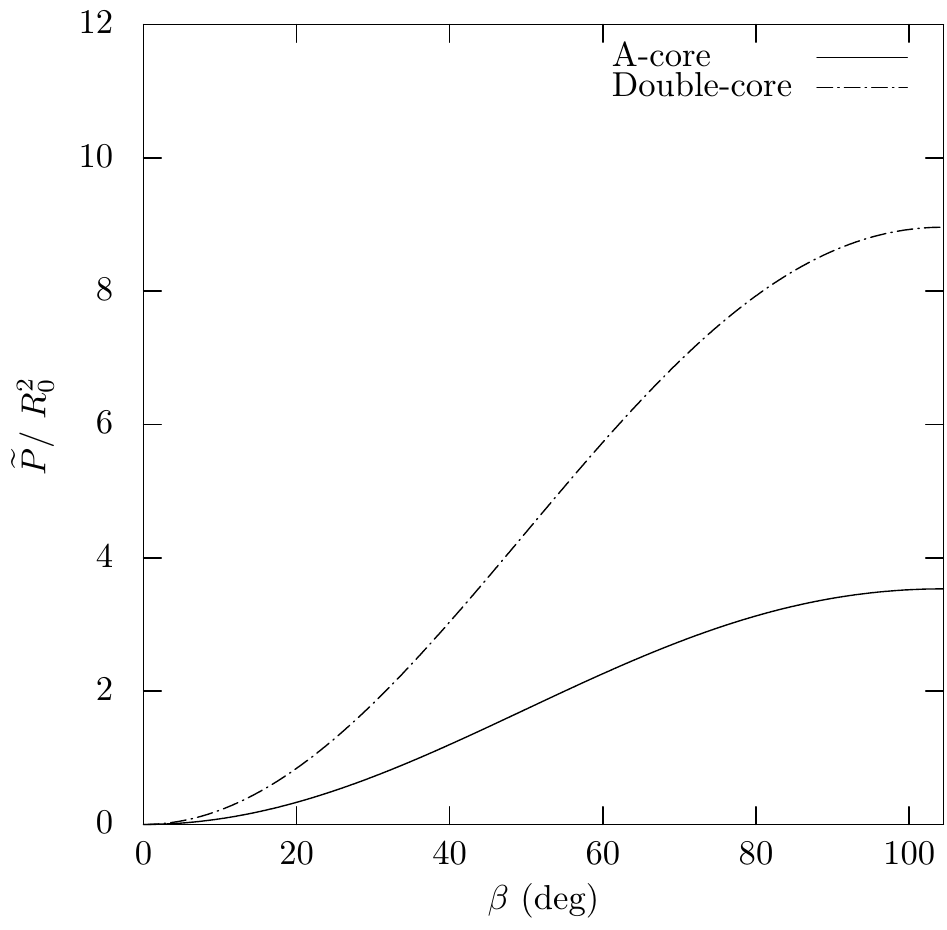}
}
\subfloat{%
\includegraphics[width=0.49\textwidth]{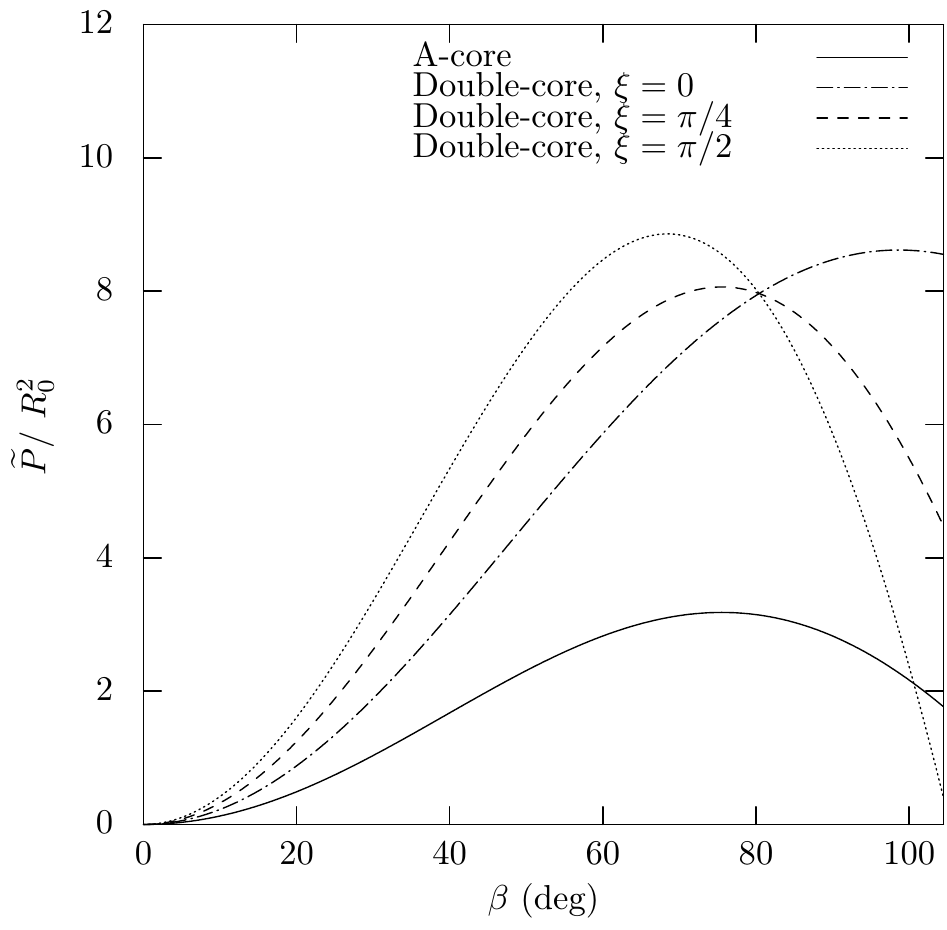}
}
\caption{$\widetilde{P} / R_0^2$ as a function of $\beta$ (in degrees) at $T = 0.5 T_c$, $p = 29.3$ bar, $c = 1$. In the left figure the magnetic field is parallel to the vortex axis ($\eta = 0$), and thus there is no dependence on $\xi$. In the right figure the magnetic field is perpendicular to the vortex axis ($\eta = \pi / 2$). In this case there is a strong dependence on $\xi$ for the double-core vortex.}
\label{fig:P_tipping}
\end{figure}

Another case to consider is to fix the field angle $\eta$ and study $\widetilde{P}$ as a function of tipping angle $\beta$. This is shown in Fig. \ref{fig:P_tipping}. In the left figure $\eta = 0$, while in the right figure $\eta = \pi / 2$. When $\eta = 0$, i.e. the magnetic field is parallel to the vortex axis, there is no dependence on the angle $\xi$. Both vortices show a similar qualitative behaviour, with $\widetilde{P}$ increasing monotonously from zero at $\beta = 0$. When $\eta = \pi / 2$, i.e. the magnetic field is perpendicular to the vortex axis, the shape of $\widetilde{P}$ is different. Again, $\widetilde{P}$ starts to increase from zero at $\beta = 0$, but now there is a maximum at an angle $\beta_{\mathrm{max}} \leq \theta_0$, which in the case of the double-core vortex depends on $\xi$.

In order to find the magnitude of the absorption, we need to estimate the cut-off radius $r_1$. A simple estimate is $r_1 = C_1$. For the double-core vortex, this is on the order of the distance between the half cores, calculated in Ref.\ \cite{silaev2015}.
An alternative prescription,  which works for the double-core vortex, is to extend the asymptotic form (\ref{eq:theta}) by the extrapolation
\begin{eqnarray}\label{eq:extrapolation}
\frac{C_1}{r}\rightarrow \arctan\frac{C_1}{r}.
\end{eqnarray}
This correctly gives that $\theta=\pi/2$ at the vortex axis $r=0$ and thus removes the need for the cut-off. As $\theta$ is no longer small, we instead of Eq.\ (\ref{eq:dipea}) use the exact the dipole energy (\ref {eq:dipegen}), and Eq.\ (\ref{eq:R2}) is replaced by 
$\langle\left| \bm R  \right|^2\rangle =  \langle|\bm R_D|^2\rangle-\langle\bm R_D\rangle\cdot\langle\bm R_D\rangle$, where 
$\bm R_D=-df_D/d\bm \vartheta$ is the dipole part of the torque. 

The coefficient $\widetilde{\tau}_{LT}$ can be obtained from  experiments. From Fig. 1(b) of Ref. \cite{bunkov1990} we extract the approximate values $\widetilde{\tau}_{LT} \approx 0.07\ \mu s$ at $T = 0.6 T_c$ and $\widetilde{\tau}_{LT} \approx 0.035\ \mu s$ at $T = 0.5 T_c$. The  numerical values of the parameters needed in Eq. (\ref{eq:P_LT}) are listed in Table \ref{tab:parameters}.

We can now calculate the magnitude of the absorption.  We study the case $\eta = 0$, $\beta = \theta_0$ and  the total vortex length is approximately $12$ m in accordance with  experiment of Ref. \cite{kondo1991}.
The calculated values of the absorption at $T = 0.5 T_c$ and $T = 0.6 T_c$ are listed in Table \ref{tab:power}. Avoiding the cut-off by the model (\ref{eq:extrapolation}), denoted by $r_1=0$, gives lower absorption than a sharp cut-off at $C_1$. Based on this we judge that the $r_1=0$ choice should be closer to the truth than  $r_1=C_1$ for the double-core vortex. This makes us suspicious that the  values for $r_1=C_1$ may be too large also for the A-phase-core vortex, although the model (\ref{eq:extrapolation}) cannot be justified there.

The last row of Table \ref{tab:power} gives the measured values of absorption \cite{kondo1991}.
We see that the theoretically calculated values are somewhat smaller, especially those for the double-core vortex which are expected to be more physical ($r_1=0$). Another problem is that the theoretical values are almost independent of temperature  (in the narrow range between $0.5 T_c$ and $0.6 T_c$) whereas the measured value changes more than by a factor of two.

\begin{table}
\renewcommand{\arraystretch}{1.5}
\begin{tabular}{| c| c | c | c | c | c | c | c | c | c |}
\hline
 & $R_0$ & $\xi_D$ &$\frac{\chi \Omega^4}{\mu_0 \gamma_0^2} \, R_0^2$ & $\widetilde{\tau}_{LT}$ & \multicolumn{2}{| c |}{Double-core} & \multicolumn{1}{| c |}{A-phase-core} \\
$T / T_c$ &  $\left( \mathrm{nm} \right)$ & $\left( \mathrm{\mu m} \right)$ & $\left( \mathrm{pW \, m^{-1} \, \mu s^{-1}} \right)$ & $\left( \mu \text{s} \right)$ & $C_1 / R_0$ & $C_2 / R_0$ & $C_1 / R_0 = C_2 / R_0$ \\
\hline
$0.5$ & $92.7$ &$9.4$ & $0.665$ & $0.035$ & $3.72$ & $0.11$ & $1.50$  \\
\hline
$0.6$ & $92.7$ &$9.3$ & $0.493$ & $0.07$ & $3.00$ & $0.08$ & $1.33$  \\
\hline
\end{tabular}
\caption{Numerical values of some relevant parameters at two different temperatures at the pressure of $29.3$ bar.}
\label{tab:parameters}
\end{table}

\begin{table}
\renewcommand{\arraystretch}{1.5}
\begin{tabular}{| c | c | c | c | c |}
\hline
& \multicolumn{2}{| c |}{$P_{\text{tot}}$ (pW), Double-core} & \multicolumn{2}{| c |}{$P_{\text{tot}}$ (pW), A-phase-core} \\
 $r_1$ & $T = 0.5 T_c$ & $T = 0.6 T_c$ & $T = 0.5 T_c$ & $T = 0.6 T_c$ \\
\hline
$C_1$ & $6.7$ & $6.9$ & $3.6$ & $4.3$  \\
\hline
$0$ & $3.0$ & $3.1$ & $-$ & $-$   \\
\hline
Experiment\ \cite{kondo1991} & $88$ & $41$ & $-$ & $15$  \\
\hline
\end{tabular}

\caption{Numerical values of the power absorption. The parameters are chosen to match the conditions of Ref. \cite{kondo1991}, with $\eta = 0$. Here $r_1 = 0$ means that we have used the extrapolation of Equation (\ref{eq:extrapolation}), which removes the need for the cut-off. Since the extrapolation model is good for the double-core vortex only, we have not used it for the A-phase-core vortex. The last row gives the measured power absorption \cite{kondo1991}.}
\label{tab:power}
\end{table}

\section{Conclusions}

We have calculated the energy dissipation rate due to the Leggett-Takagi relaxation using analytic expression for the order parameter outside the vortex core.
There is qualitative and also some quantitative agreement with the experiment. However, the calculated numerical values for the double-core vortex are smaller than the measured ones. Part of this could be due to the uncertainties in the values of the relaxation time, the cut-off radius and $C_1$, but it seems unlikely that these could account for the temperature dependence of the absorption between $T=0.5 T_c$ and $T=0.6 T_c$.
We expect that an important role is played by spin dynamics, which should lead to additional dissipation in the form of radiation of spin waves. This will be studied in a separate publication.


\begin{acknowledgements}
We thank the Academy of Finland and Tauno T\"onning foundation for financial support.
\end{acknowledgements}

\pagebreak

\end{document}